\documentclass[anonymous,runningheads]{llncs}
\usepackage{graphicx, cite, subfiles, booktabs, diagbox, amssymb, threeparttable,adjustbox,}
\usepackage{amsmath} 
\usepackage{orcidlink}

\usepackage{array}
\newcolumntype{C}[1]{>{\centering\arraybackslash}m{#1}}

\begin{document}
\title{Multimodal Deformable Image Registration for Long-COVID Analysis Based on Progressive Alignment and Multi-perspective Loss}

\authorrunning{J. Li et al.}
\titlerunning{CT-MRI Progressive Registration Based on MPL}

\author{Jiahua Li\inst{1,2}\orcidlink{0009-0009-7903-9875} \and James T. Grist\inst{3}\orcidlink{0000-0001-7223-4031} \and Fergus V. Gleeson\inst{3}\orcidlink{0000-0002-5121-3917} \and Bart\l omiej W. Papie\.z\inst{1,2}\orcidlink{0000-0002-8432-2511}}

\institute{
$^{1}$Nuffield Department of Population Health, University of Oxford, UK \\
$^{2}$Big Data Institute, University of Oxford, UK \\
$^{3}$ Oxford Radiology Research Unit, Oxford University Hospitals NHS Foundation Trust, UK \\
\email{\{jiahua.li@ndph, bartlomiej.papiez@bdi\}.ox.ac.uk}
}
\titlerunning{CT-MRI Progressive Registration Based on MPL}
% \author{}
% \institute{}

\maketitle

\begin{abstract}
Long COVID is characterized by persistent symptoms, particularly pulmonary impairment, which necessitates advanced imaging for accurate diagnosis. Hyperpolarised Xenon-129 MRI (XeMRI) offers a promising avenue by visualising lung ventilation, perfusion, as well as gas transfer. Integrating functional data from XeMRI with structural data from Computed Tomography (CT) is crucial for comprehensive analysis and effective treatment strategies in long COVID, requiring precise data alignment from those complementary imaging modalities. To this end, CT-MRI registration is an essential intermediate step, given the significant challenges posed by the direct alignment of CT and Xe-MRI. Therefore, we proposed an end-to-end multimodal deformable image registration method that achieves superior performance for aligning long-COVID lung CT and proton density MRI (pMRI) data. Moreover, our method incorporates a novel Multi-perspective Loss (MPL) function, enhancing state-of-the-art deep learning methods for monomodal registration by making them adaptable for multimodal tasks. The registration results achieve a Dice coefficient score of 0.913, indicating a substantial improvement over the state-of-the-art multimodal image registration techniques. Since the XeMRI and pMRI images are acquired in the same sessions and can be roughly aligned, our results facilitate subsequent registration between XeMRI and CT, thereby potentially enhancing clinical decision-making for long COVID management.
\end{abstract}


\section{Introduction}
Considering the thorough documentation of over 651 million COVID-19 cases worldwide, the current conservative estimates suggest that around 65 million people are suffering from long COVID \cite{ballering2022persistence}.
What is more, a number of patients with long COVID present no findings in Computed Tomography (CT), and more advanced imaging techniques such as hyperpolarized Xenon MRI (XeMRI) have to be utilised to detect lung abnormalities \cite{grist2022lung}. 
While XeMRI provides insight about the lung function, it needs to be analysed with respect to the underlying anatomy (shown e.g. in CT) to be utlised in clinical decision-making consequently requiring multimodal image registration for this task.

Monomodal deformable image registration (DIR) is regarded as a non-trivial task, due to patient motion \cite{papiez2014implicit, hua2017multiresolution, de2019deep, anas2020ct} (for longitudinal studies) or the subject variability \cite{ehrhardt2010statistical} (for cross-sectional studies). Nevertheless, the complexity of DIR increases in the multimodal scenarios, fueled by differences in intensities between the images acquired to visualise diverse physical phenomena, e.g. CT or MRI, where each relies upon different physical properties of tissue to create images. Since multimodal DIR gives clinicians more comprehensive insights about a patient's condition, benefiting diagnostic accuracy and personalised treatment plans, efficient multimodal DIR is critical, and many methods have been suggested \cite{sotiras2013deformable}. However, statistical and information theory-based methods suffer from computational complexity and slow convergence \cite{wells1996multi,maes1997multimodality,hermosillo2002variational}, while descriptor-based methods prove sensitive to initial conditions and require effective pre-alignment to handle extensive translations\cite{heinrich2012mind,heinrich2013towards}; their reliance on hand-crafted features calls for domain expertise for fine-tuning and restricts their adaptability.
As of late, Convolutional Neural Networks (CNNs) have been utilised to learn a standard representation for DIR by optimising a similarity metric \cite{hu2018label, hu2018weakly,guo2019multi}. Concurrently, selecting appropriate similarity metrics proves challenging since multimodal images can exhibit differences in intrinsic intensity distribution and resolution, leading to the effectiveness of learning-based methods being limited mainly in the monomodal scenarios \cite{balakrishnan2019voxelmorph,zhao2019recursive,zheng2022recursive, ZHENG2024103038}. Alternatively, multimodal DIR can be transformed into a less complex monomodal task utilising an image-to-image (I2I) translation \cite{qin2019unsupervised}. %\cite{qin2019unsupervised,chen2022unsupervised}. 
Nonetheless, such translation can potentially result in shape inconsistency and produce artificial anatomical features, further deteriorating the performance of the DIR. 

The focus of this work is on the DIR between CT and proton MRI (pMRI), a process of significance to the analysis of XeMRI. Owing to its non-ionising characteristics, XeMRI has gained considerable interest for long COVID, primarily due to capturing images related to lung ventilation, perfusion, and gas transfer in lungs\cite{albert1994biological,mugler2013hyperpolarized,szmul2019patch, grist2022lung}. Since XeMRI does not provide anatomical information, the alignment of XeMRI images with pMRI and CT is essential.
pMRI is typically acquired in the same imaging session as XeMRI, albeit not within the same breath-hold, while CT is taken a couple of days prior. This poses a challenge when attempting to fuse XeMRI with CT, thus necessitating DIR between pMRI and CT.

Contributions of our work are as follows. To overcome the aforementioned limitations, we proposed a multimodal, end-to-end method based on progressive alignment architecture which can tackle significant deformations (Sec.~\ref{subsec:paa})). Moreover, we introduce a novel Multi-perspective Loss (MPL) function, applicable to any existing monomodal DIR architecture, extending their application to multimodal imaging registration (Sec.~\ref{subsec:mpl}).
Lastly, our method was evaluated on challenging long-COVID lung CT and pMRI dataset, which achieved the Dice coefficient (DSC) of 0.91, outperforming the state-of-the-art models for multimodal DIR (Sec.~\ref{sec:exp}). To the best of our knowledge, this is the first effort to automate mutlimodal deformable image registration for long-COVID CT and pMRI.

\section{Methodology}
\subsection{Overview}
As seen in Fig.~\ref{fig:proposed_architecture}, DIR aims to estimate a non-linear voxel-to-voxel correspondence between a fixed image $F$ and a moving image $M$, in which the estimated transformation is parameterized with $\phi$: \begin{equation}
    \phi = f_{\theta}(M, F)
\end{equation}
with $f$ and $\theta$ corresponding to the utilised neural networks and the networks' learning parameters, respectively. Our method uses two 3D images as input: the pMRI image (the moving image) and the CT image serving as a reference (the fixed image). These are introduced into a cascading sequence of 3D CNNs (described in Sec. \ref{subsec:paa}) to extract distinctive feature maps from both input images. Furthermore, we use a novel loss function (described in Sec. \ref{subsec:mpl}) that combines Mutual Information (MI) and Gaussian Pyramid labels to capture both global and local intensity information. 
In this section, the workflow of our methodology is outlined, with detailed description in the subsequent subsections.

\subsection{Progressive Alignment Architecture}
\label{subsec:paa}
\begin{figure*}[t]
  \centering
  \includegraphics[width=1.0\textwidth]{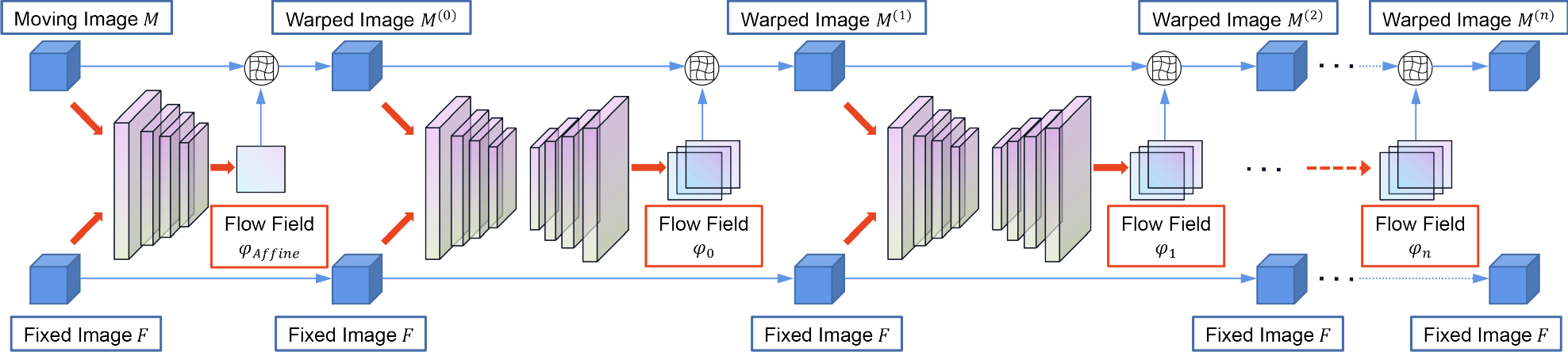}
  \caption{Illustration of our proposed architecture: The network initially predicts the affine transformation $\phi_{\text{affine}}$. Subsequently, cascaded networks iteratively compute the displacement fields. At each cascade $n$, the network predicts the displacement field $\phi_{\text{n}}$ using the input image $M^{\text{(n)}}$ warped by the displacement field $\phi_{\text{n-1}}$ from the previous cascade $n-1$.}
  \label{fig:proposed_architecture}
\end{figure*}

As a consequence of the significant deformation observed across diverse modalities, estimation of the displacement field in one attempt proves to be challenging. Thus, the model is implemented iteratively to ensure progressive refinement (see~Fig.\ref{fig:proposed_architecture}). The first iteration aims to establish the coarse transformation, while subsequent refinements estimate finer transformations. Specifically, the suggested model is initiated by a network that predicts an affine transformation matrix with 12 degrees of freedom (denoted by $\phi_{\text{affine}}$ in Fig.~\ref{fig:proposed_architecture}). The network for affine transformation has four downsampling residual-network blocks (ResBlock).
%and the initial channels of four. 
The final convolutional layer employs a fully-connected matrix, subject to learning, to create a linear projection, producing a vector encompassing 12 parameters for affine transformation. Following the network for the affine alignment, cascades of registration networks (sharing weights) predicting dense displacement fields (DDF) are employed to estimate local (non-rigid) deformation~$\phi_{n}$.
Similarly, each cascaded network has a Voxelmorph-style architecture\cite{balakrishnan2019voxelmorph}, replacing the encoder component with four downsampling ResBlocks. The affine transformation and the DDFs are recursively estimated by the Spatial Transformer Network (STN) \cite{jaderberg2015spatial} to produce the final DDF $\phi$. 
Considering the $n$-th cascade, the output will be estimated according to the DDF $\phi_{n-1}$ from the $(n-1)$-th cascade: \begin{equation}
    f_{\theta}(M, F) = (\phi_{n-1} \circ \phi_{n}) + \phi_{n}
\end{equation} where $\circ$ corresponds to the warping operation facilitated by a trilinear image resampler. Theoretically, this recursive process can be infinitely applied. Hence, the input image $M$ becomes warped by the final DDF $\phi$ estimated according to its affine transformation and multiple cascades of deformable transformation, resulting in the registered image $M^{\prime}$, represented as: \begin{equation}
M^{\prime} = \phi \circ M
\end{equation}

\subsection{Multi-perspective Loss}
\label{subsec:mpl}

The deformations observed in the thorax and complementary information captured by both pMRI and CT require the model to pay attention to not only the local information (like edges, textures, and corners) but long-range (dis)similarities. 
To address the challenge, we proposed a novel loss function: the multi-perspective loss (MPL), including the Mutual Information (MI) and Gaussian Pyramid label (GPL) loss. The MI loss,
\begin{equation}
MI_{\text{loss}}(M, F) = -\sum_{m \in M} \sum_{f \in F} p(m, f) \log \left( \frac{p(m, f)}{p(m)p(f)} \right),
\end{equation}quantifying the statistical discrepancy between two images, focuses on global alignment. Simultaneously, to overcome its limitations for local alignment, the GPL loss is employed, which used Gaussian filters,\begin{equation}
G(x, y, z) = \frac{1}{(2\pi\sigma^2)^{3/2}} \exp\left( -\frac{x^2 + y^2 + z^2}{2\sigma^2} \right),\end{equation}
to derive feature pyramids across various scales, thereby facilitating the capture of local correspondences between images. Specifically, segmentation labels from MRI $M_{label}$ and CT $F_{label}$ images were selectively filtered by 3D Gaussian kernels, operating at six separate standard deviation scales, \(\sigma \in \{0, 1, 2, 4, 8, 16\}\). The higher scales encourage the model to focus on the entire lung cavity, while the lower scales target more local features, such as edges and corners. This dual focus enables the alignment of both large-scale structural features and smaller, more intricate details, thus addressing the MI loss's limitation of neglecting anatomical information. As such, the resulting loss function is denoted as follows: \begin{equation}
\begin{aligned}
    L(M,F,\phi) &= \alpha L_{MI}(M,F,\phi) \\
                &\hspace{0.0em} + \beta 
                L_{GPL}(M_{label},F_{label},\phi)
                 \\
                &\hspace{0.0em} + \lambda L_{reg}(\phi)
\end{aligned}
\label{eq:loss_function}
\end{equation}where $L_{MI}$ and $L_{GPL}$ represent the MI and GPL losses. The function $L_{reg}$ is a regularisation term using a weighted bending energy \cite{rueckert1999nonrigid} to penalise local spatial variations in $\phi$, ensuring a smooth displacement field. The parameters $\alpha$, $\beta$ and $\lambda$ serve as the weighting coefficients, modulating the contribution of every corresponding term in the loss function, respectively.

\section{The Experiment}
\label{sec:exp}
\subsection{Dataset}
We conducted an assessment of the proposed method using an in-house Post-COVID Assessment Clinic dataset, including 46 pairs of CT, pMRI and XeMRI images. Specifically, MRI was performed at 3 T (GE Healthcare, Premier) using a phased array thoracic imaging coil (30 channels). Proton imaging consisted of a 3D spoiled gradient echo sequence, characterized by a Repetition Time (TR) of 3.1 ms, Echo Time (TE) of 1 ms, Field of View (FOV) of 400 mm, slice thickness of 5 mm, an acquisition matrix of \(256 \times 128\), a reconstruction matrix of \(256 \times 256\), number of slices = 36, performed in a single breath-hold, with a bandwidth of 62.5 kHz and a flip angle of 20 degrees. 

Subsequent to inhalation of 1L of polarized Xenon-129, XeMRI was acquired using a Transmit/Receive vest coil (PulseTeq, Cobham, UK) employing a 4-echo radial sequence with TR = 23 ms, an acquisition matrix of \(16 \times 16 \times 16\), a reconstruction matrix of \(32 \times 32 \times 32\), FOV of 400 mm, a flip angle of 40 degrees, and using Iterative Decomposition of Water and Fat with Shifted Echo Times and Lease Squares Regression (IDEAL) Reconstruction.

CT was performed using a GE Healthcare system with a section thickness of 0.625 mm and a slice resolution of \(512 \times 512\) after an inhalation of 1L of room air. 

All images were resampled as isotropic, with a spatial resolution of \(5 \times 5 \times 5\,\text{mm}^3\). Subsequently, the images were cropped based on the lung region, followed by padding to the size of \(128 \times 128 \times 128\). The dataset was then randomly split into 30 pairs for training, 6 for validation, and 10 for testing.
All reported results presented within this study are derived from the analysis conducted on the testing dataset.

\subsection{Implementation Details}
Our method was implemented using Pytorch on an NVIDIA RTX6000 GPU. All models were trained for 300 epochs, with a batch size of 1 and the experiments of five-fold cross-validation. To ensure the most favourable results, we set cascades in our method to 5. The Adam optimiser was utilised, with a \(1 \times 10^{-5}\) learning rate. Lastly, hyperparameters for our loss function, $\alpha$, $\beta$ and $\lambda$ are set to 1.0, 1.0 and 2.0. These values were carefully optimized to achieve a balanced improvement in training stability, registration accuracy, and transformation invertibility.

\subsection{Comparison with the state-of-the-art methods}
The proposed model was benchmarked against the state-of-the-art iterative DIR: SyN \cite{avants2008symmetric,szmul2018xemri}, and deep learning based DIR methods: VXM \cite{balakrishnan2019voxelmorph}, RCN \cite{zhao2019recursive}, and CompositeNet (CompNet)\cite{hu2018weakly}. First, SyN was implemented using ANTsPy. Next, VXM used a U-Net for non-iterative registration, while RCN employed an iterative approach with a Volume Tweening Network (VTN) configuration \cite{zhao2019recursive}. Initially, VXM and RCN adopted Normalised Cross Correlation (NCC) loss and $L2$ variation loss as a regularisation. However, since the NCC loss leads to poor registration results, we further substituted the NCC loss with the proposed MPL (See Eq. \ref{eq:loss_function}). This comparison was conducted to underscore the superior applicability of the suggested loss across the state-of-the-art models. Furthermore, CompNet is a popular multimodal DIR method consisting of the GlobalNet and the LocalNet. As such, it encourages both global and local alignment, calculating the loss by seven scales of the DSC and the weighted bending energy as the regularisation \cite{hu2018weakly,rueckert1999nonrigid}. Registration accuracy was evaluated by measuring the overlap between registered and fixed segmentation masks with the Dice Similarity Coefficient (DSC) \cite{dice1945measures}. The percentage of negative Jacobian determinants on the estimated displacement fields (\(\%J_{\phi}\)) allowed for a further assessment of the transformation invertibility with a lower \(\%J_{\phi}\) indicating smoother transformations. 
The traditional methods are evaluated on the same testing data, while all the state-of-the-art deep learning-based methods are trained and tested on the same splits of the dataset.

% Table 1

% \begin{table}[h] %h for here
% \begin{table}[b]  %b for bottom
\begin{table}[t]  %t for top
\renewcommand{\arraystretch}{1.2}
\centering
\caption{Quantitative evaluation results of the proposed and comparison methods.}
% \begin{adjustbox}{width=0.4\textwidth}
\begin{tabular}{|C{1.3cm}|C{3.1cm}|C{1cm}|C{1cm}|}
     \hline
     Methods & Loss Function & DSC & \(\%J_{\phi}\)\\
     \specialrule{1.5pt}{0pt}{0pt}
     Initial & \diagbox{}{} & 0.671 &  \diagbox{}{}\\
     \hline
     SyN & MI & 0.693 & \diagbox{}{}\\
     \hline
     VXM & NCC + Dice Loss & 0.691 &  \textbf{0.31}\% \\
     \hline
     VXM & MIND & 0.701 & 0.35\%
     \\
     \hline
     VXM & MPL & 0.789 & 0.51\% \\
     \hline
     RCN & NCC + Dice Loss & 0.695 & 0.49\% \\
     \hline
     RCN & MPL & 0.895 & 1.89\% \\
     \hline
     CompNet & Multi-scale Dice Loss& 0.848 & 0.76\% \\
     \hline
     Ours& MPL & \textbf{0.913} & 0.89\% \\
     \hline
\end{tabular}
% \end{adjustbox}
\label{tab:compare}
\end{table}

\section{Results and Discussion}
\subsection{Registration Results}
% Fig. 2
\begin{figure*}[htb]
  \centering
    \includegraphics[width=1.0\textwidth]{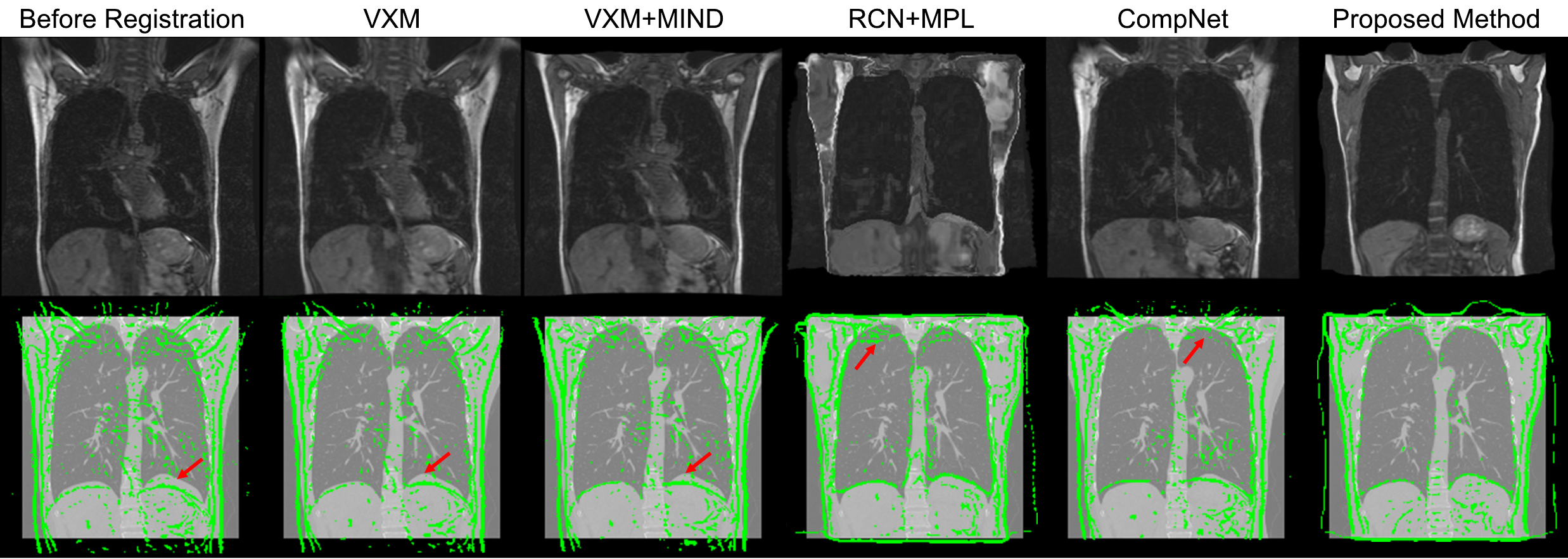}
  \caption{Visualisation of CT-pMRI registration through a wide range of methods: The top row reflects the input pMRI image before registration and the results from state-of-the-art methods, whereas the bottom row shows the fixed image (CT) overlaid by edges extracted from pMRI image (in green). Red arrows point out areas of mis-registration.}
  \label{fig:vis}
\end{figure*}
Results are summarised in Tab.~\ref{tab:compare} and visualised in Fig.~\ref{fig:vis}. Our model demonstrates superior performance compared to state-of-the-art methods, achieving the highest Dice Similarity Coefficient (DSC) of 0.91, in contrast to the second-best performing method (CompNet), which attained a DSC of 0.84 only. Even though the \(\%J_{\phi}\) is 0.89 \% for our method (comparing to 0.31\% for VoxelMorph), it remains situated within an acceptable range, pointing to sufficient transformation invertibility. Intuitively, the results of VXM and RCN, with the NCC loss, point out a marginal improvement in registration performance. Nevertheless, integrating the MPL considerably boost to their registration accuracy, emphasising the robustness of the proposed registration loss. Thus, our loss has the potential to enhance the performance of any existing state-of-the-art models tailored to monomodal scenarios and enable them to address challenging multimodal image registration like pMRI and CT outlined in this paper.

% Table 2

\begin{table}[t] %h for here
% \begin{table}[b]  %b for bottom
\renewcommand{\arraystretch}{1.2}
\centering
\caption{Comparison of the proposed model with varied cascade configurations.}
% \begin{adjustbox}{width=0.4\textwidth}
\begin{tabular}{|C{3cm}|C{2cm}|} 
     \hline
     Methods &  DSC \\
     \specialrule{1.5pt}{0pt}{0pt}
     Initial & 0.671 \\
     \hline
     1 cascade & 0.871 \\
     \hline
     2 cascades & 0.882  \\
     \hline
     3 cascades & 0.896  \\
     \hline
     4 cascades & 0.902 \\
     \hline
     5 cascades & \textbf{0.913} \\   
     \hline
\end{tabular}
% \end{adjustbox}
\label{tab:cascade}
\end{table}

With the aim of further assessing the effectiveness of our method, an evaluation was conducted using different configurations of cascades. The configurations varied in the number of cascades exhaustively detailed in Tab.~\ref{tab:cascade}. A systematic approach was adopted to explore the effect of each configuration on the method's general performance. Accordingly, the results showcase that the architecture incorporating five cascades achieves the highest registration accuracy within the evaluated range. While architectures with more than five cascades lacked exploration due to computational limits,  the timely findings are firmly in favour of the efficacy of the five-cascade design.

\subsection{Ablation Study}
Our novel loss function combines the advantages of the MI and the multi-scale label loss, enhancing the accuracy of multimodal DIR. As seen in  Tab.~\ref{tab:ablation}, an ablation study assesses every component's impact of each similarity measure in our loss function. In addition, we compare our method to the most relevant method i.e. CompNet by Hu et al.\cite{hu2018weakly}.
The results indicate that by combining global and local information, our method can efficiently register challenging multimodal images such as pMRI and CT.

% Table 3

\begin{table}[t]
\renewcommand{\arraystretch}{1.2}
\centering
\caption{The ablation study results on: 1) Mutual Information (MI) loss 2) Gaussian-pyramid label loss}
\begin{threeparttable}
\begin{tabular}
{|C{3cm}|C{1cm}|C{1cm}|C{2cm}|}
     \hline
     Loss Functions & 1) & 2) & DSC \\
     \specialrule{1.5pt}{0pt}{0pt}
     % \hline
     Initial & \diagbox{}{} & \diagbox{}{} & 0.671 \\
     \hline
     Hu et al. \cite{hu2018weakly} \tnote{*} & $\times$   &  $\checkmark$  & 0.879 \\
     \hline
     MI  & $\checkmark$  &  $\times$ & 0.760 \\
     \hline
     Gaussian-pyramid & $\times$ & $\checkmark$ & 0.890 \\
     \hline
     Multi-perspective & $\checkmark$ & $\checkmark$ & \textbf{0.913} \\ 
     \hline
\end{tabular}
\begin{tablenotes}[para,flushleft]
\footnotesize
    \item[*] We employ the loss function proposed by Hu et al., applying it to our proposed network instead of CompNet.
\end{tablenotes}

\label{tab:ablation}
\end{threeparttable}
\end{table}

\subsection{CT-XeMRI Registration}
\begin{figure*}[t]
  \centering
  \includegraphics[width=0.5\textwidth]{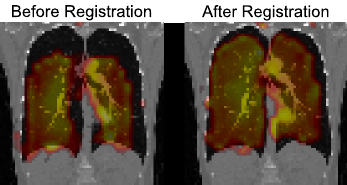}
  \caption{Visualization of the CT-XeMRI results before and after registration.}
  \label{fig:xe_ct_reg}
\end{figure*}

The pMRI and XeMRI images are acquired within the same session, ensuring inherent alignment. Utilizing the transformation matrices derived from the CT-pMRI registration via our proposed network, we can facilitate the CT-XeMRI registration, as illustrated in Fig.~\ref{fig:xe_ct_reg}. This process aligns the structural and functional data, which is instrumental in clinical analyses that explore the relationship between anatomical and functional impairments. However, the acquisition of pMRI and XeMRI images during distinct breath-hold intervals introduces some degree of misalignment. Future research will aim at addressing this breath-hold variability to enhance the pMRI-XeMRI alignment, thereby improving the precision of CT-XeMRI registration.

\section{Conclusion}
This paper presented an end-to-end model based on progressive alignment for multimodal DIR. Our novel loss function enhances the performance of cutting-edge models formerly restricted to monomodal scenarios, promoting their utilisation in multimodal imaging registration scenarios. The proposed methods outperformed existing ones when evaluated on challenging 3D lung images from CT and pMRI. Notably, this work can significantly advance multimodal image analysis, offering a pivotal contribution that holds the potential to reshape our understanding and method for long-COVID research.

\section{COMPLIANCE WITH ETHICAL STANDARDS}
This study was performed in line with the principles of the Declaration of Helsinki. Approval was granted by the South Central - Oxford C Research Ethics Committee on 15 Dec 2021 (reference 21/SC/0398).

\section{ACKNOWLEDGEMENTS}
This study is funded by the National Institute for Health and Care Research (NIHR) (Long Covid grant, Ref: COV‐LT2‐0049). The views expressed in this publication are those of the authors and not necessarily those of NIHR or The Department of Health and Social Care. 

\bibliographystyle{splncs04}
\bibliography{ref}

\end{document}